\documentclass{article}

\usepackage{PRIMEarxiv}

\usepackage[utf8]{inputenc} 
\usepackage[T1]{fontenc}    
\usepackage{hyperref}       
\usepackage{url}            
\usepackage{booktabs}       
\usepackage{amsfonts}       
\usepackage{nicefrac}       
\usepackage{microtype}      
\usepackage{lipsum}
\usepackage{amsmath}
\usepackage{graphicx}
\graphicspath{{media/}}     
\usepackage{graphicx}
\usepackage{algorithm}      
\usepackage{algpseudocode} 
\usepackage{mdframed}
\usepackage{mdframed}
\usepackage{mathtools}

\usepackage{amsmath}
\usepackage{subfigure}
\usepackage{multicol}
\usepackage{multirow}
\usepackage{siunitx}
\usepackage{dirtytalk}
\usepackage{hyperref}
\usepackage{arydshln}
\usepackage{xcolor}
\usepackage{mathtools}
\usepackage{tikz}
\usepackage{afterpage}
\usepackage{caption}
\usepackage{adjustbox}
\usepackage{array}
\usepackage{ragged2e}
\usepackage{bbding}
\usepackage{setspace}   
\usepackage[switch]{lineno}

\usepackage{amsmath}
\usepackage{breqn}

\def\etal{et al.}

\title{Adversarial Distortion Learning for Medical Image Denoising
}

\author{
  Morteza Ghahremani \\
   Technical University of Munich \\
  \AND
  Mohammad Khateri \\
  University of Eastern Finland \\
  \And
  Alejandra Sierra \\
  University of Eastern Finland \\
  \And
  Jussi Tohka \\
  University of Eastern Finland \\
}

\begin{document}
\maketitle

\begin{abstract}
We present a novel adversarial distortion learning (ADL) for denoising both two- and three-dimensional (2D/3D) biomedical image data. 
The proposed ADL methodology deploys two interconnected auto-encoders, a denoiser and a discriminator. 
The primary function of the denoiser is to effectively eliminate noise from the input data, while the discriminator plays a crucial role in assessing and comparing the denoised image against its noise-free counterpart. 
This iterative process is perpetuated until the discriminator attains the inability to differentiate between the denoised data and the corresponding reference image. 
The foundational elements of both the denoiser and the discriminator are founded upon the proposed Efficient-UNet. 
The Efficient-UNet encompasses a novel multiscale architecture, effectively utilizing residual blocks and introducing a novel pyramidal approach in the backbone to facilitate the efficient extraction of texture and the reuse of feature maps. By virtue of this innovative architectural design, the model ensures the coherence of the denoised data concerning both its local and global attributes. 
Traditional adversarial learning methods often fall short in preserving crucial textural, global, and local data representations. Our proposed Efficient-UNet is equipped with two novel loss functions to conserve the inherent textural information and contrast during the training phase to overcome this limitation. 
We generalized the proposed ADL to any biomedical data. 
The 2D version of our network was trained on natural images  
and tested on biomedical datasets featuring entirely distinct distributions from natural images, so re-training is unnecessary.
Experiments on magnetic resonance images (MRI), dermatoscopy, electron microscopy, and X-ray datasets show that the proposed method achieved state-of-the-art denoising each benchmark. 
Our implementation and pre-trained models are available at~\url{https://github.com/mogvision/ADL}.
\end{abstract}


\section{Introduction}\label{intro}\sloppy
\doublespacing

Image denoising aims to recover latent noise-free two/three-dimensional (2D/3D) image data $\mathbf{x}$ from its noisy counterpart $\mathbf{y}=\mathbf{x}+\mathbf{n}$, where $\mathbf{n}$ denotes noise that is independent of
$\mathbf{x}$.
The denoising problem is an ill-posed inverse problem as no unique solution exists. 
Denoising is an essential preprocessing step in many medical imaging applications. A vast number of methods have been proposed over the past decades~\cite{gal2009denoising,manjon2010adaptive,kaur2018review,sagheer2020review}, and recently, methods based on deep neural networks (DNNs) attract more attention due to their good performance~\cite{gondara2016medical,yang2018low,sharif2020learning,ma2020cycle,zhang2021plug}. 

For additive noise, it is possible to estimate denoised visual data$\mathbf{x}$ using the Bayesian approach, where the posterior $P_{x|y}(\mathbf{x}|\mathbf{y})$ is a combination of the data likelihood $P_{y|x}(\mathbf{y}|\mathbf{x})$ and a prior model $P_{x}(\mathbf{x})$, i.e.:
\begin{equation}
\mathbf{\hat x} = \operatorname*{argmax}_{\mathbf{x}}\;P_{x|y}(\mathbf{x}|\mathbf{y}) 
= \operatorname*{argmax}_{\mathbf{x}}\;\text{log}\:P_{y|x}(\mathbf{y}|\mathbf{x}) +\text{log}\:P_{\mathbf{x}}(\mathbf{x}).
\label{eq:denoising1}
\end{equation}
The denoising problem Eq.~\ref{eq:denoising1} can also be expressed as an optimization of a data 
term penalized by one or more regularization terms as follows:
\begin{equation}
\mathbf{\hat x} = \operatorname*{argmin}_{\mathbf{x}} \:\left\Vert\mathbf{y}-\mathbf{x}\right\Vert^p_p+\lambda \mathcal R(\mathbf{x}),
\label{eq:denoising2}
\end{equation}
where $\left\Vert.\right\Vert_p$ denotes the $\ell{p}$-norm, $1\leq p\leq2$. $\mathcal R$ poses penalty terms on the unknown latent $\mathbf{x}$ which is associated with the prior term $P_{x}(\mathbf{x})$ defined in Eq.~\ref{eq:denoising1}. 
$\lambda$ is a Lagrangian parameter, which can be determined manually or automatically~\cite{tiddeman2021principal}. 
In general, the denoising methods can be divided into model- and learning-based categories. The model-based approach solves Eq.~\ref{eq:denoising2} using one or several regularization terms~\cite{getreuer2012rudin,mairal2009non,maggioni2012nonlocal,zhang2016statistical}. In the learning-based approach, a model can learn features with supervision~\cite{elad2006image,papyan2017convolutional} or it may learn the features simultaneously during image reconstruction~\cite{rai2021unsupervised}. 

Recently, the deep learning-based approaches~\cite{gondara2016medical,sharif2020learning,zhang2017beyond,liang2021swinir, zhu2017unpaired,bera2021noise,lee2021iscl,liu2018applications,li2021assessing} have reached excellent performance in denoising biomedical images by training DNNs on paired or unpaired images of target clean and noisy data. 
Despite the success of the existing denoising techniques, images denoised 
by such methods still suffer from poor preservation of texture and contrast. 
Other limitations of the current techniques are requiring prior knowledge of noise and the lack of generalizability to various types of biomedical image data.

In this study, we introduce an innovative adversarial distortion learning-based (ADL) method that does not require prior knowledge about the image noise.
ADL is a supervised feature learning method comprising a denoiser and a discriminator that mutually optimizes each other to find high-quality denoised contents. The denoiser and the discriminator are built upon a novel network called Efficient-UNet. Efficient-UNet has a light architecture with a pyramidal residual blocks backbone that enforces the higher-level features align with each other hierarchically.  
A remarkable feature of our approach is that it can denoise any 2D/3D biomedical image data without re-training. The light architecture of the proposed method allows a fast evaluation of test data, considerably reducing the computational time of the conventional model-based approaches. 
The key contributions of this study are as follows:
\begin{itemize}
    \item We design a novel pyramidal learning scheme, named Efficient-UNet, to further participate high-level features in the output results during training. Efficient-UNet does not need training data whose distribution is close to that of the test data, so improving the generalizability of the proposed network.
    \item We introduce a novel pyramidal loss function using `algorithme à trous' (ATW)~\cite{starck2007undecimated}. The proposed loss function keeps textural information of the reference data without amplifying noise's side effects in non-textured regions. 
    \item To preserve the histograms of reference images, we propose a novel histogram-based loss function for improving the appearance contents of denoised images.
    \item We provide both 2D and 3D networks of our method for denoising any type 
    of 2D/3D biomedical image data.
\end{itemize}
The experimental outcomes demonstrate that the proposed ADL outperforms existing methods, attaining state-of-the-art results across all benchmarks. Moreover, ADL effectively addresses issues related to overfitting, ensures generalizability to diverse biomedical data types, and mitigates computational burdens.
\section{Related Work}
Medical image denoising approaches can be grouped into two subcategories: model-based and learning-based methods. 
In the model-based approach, the optimization of Eq.\ref{eq:denoising1} solely based on the likelihood term is typically ill-posed. 
To address this issue and stabilize the denoised outputs, one or more regularization terms are incorporated alongside the data fidelity term. As a result, a diverse range of model-based techniques has been developed, encompassing total variation (TV) regularizers\cite{getreuer2012rudin, zhang2016statistical,sidky2008image} and non-local self-similarity regularizers~\cite{mairal2009non, maggioni2012nonlocal,kong2017new,chen2019denoising}, among others.
TV-based regularization terms can successfully recover piecewise constant images but cause several artifacts to complex images with rich edges and textures. 
Since images tend to contain repetitive edge and textural information, a combination of non-local self-similarity~\cite{mairal2009non,maggioni2012nonlocal,kong2017new} with the sparse representation~\cite{papyan2017convolutional} and low-rank approximation~\cite{cai2014cine} lead to significant improvements over their local counterparts. 
Despite the acceptable performance of the model-based methods in image denoising, there are still several drawbacks with these techniques. Requiring a specific model for a single denoising task, lack of generalizability to various types of data, or the need for manually or semi-automatically tuning parameters are the challenges that are still required to be addressed. Moreover, the non-local self-similarity-based methods iteratively optimize Eq.~\ref{eq:denoising2} resulting in slow convergence. 

Learning-based methods aim to learn a model's parameters with available training data. Sparsity-based techniques are well-studied learning approaches, which represent local image structures with a few elemental structures, so-called atoms from an off-the-shelf transformation matrix-like Wavelets~\cite{guleryuz2007weighted} or a learned dictionary~\cite{xu2012low,rai2021unsupervised}. 
Deep neural networks (DNNs) have been widely used for the enhancement of biomedical images, ranging from 
magnetic resonance imaging (MRI)~\cite{liu2018applications,stimpel2019multi}, computed tomography (CT)~\cite{yang2018low,wu2020self}, X-ray~\cite{gondara2016medical}, to electron microscopy (EM)~\cite{quan2019removing,lee2021iscl}. 
The learning process of DNNs can be categorized into supervised~\cite{zhang2017beyond,sharif2020learning,liang2021swinir} or unsupervised~\cite{zhu2017unpaired,bera2021noise,lee2021iscl} approaches. 
Supervised learning DNNs consider clean and noisy image pairs for training where the noisy counterparts are obtained by adding synthesized noise to the target clean ones. 
To address the challenge of insufficient clean data for training, unsupervised methods~\cite{zhu2017unpaired,huang2020noise,lee2021iscl} have been developed that estimate the map of noise from unpaired images, leveraging the supervision of clean targets. 
Unsupervised techniques often explore the noise map by generative adversarial networks (GANs)~\cite{goodfellow2014generative} and its variants like conditional GAN~\cite{mirza2014conditional} or CycleGAN~\cite{zhu2017unpaired}. Typically, the supervised DNNs have shown superior performance over the unsupervised and conventional model-based approaches. 
Despite the great success of DNNs in denoising biomedical images, 
denoised images still suffer from poor textural information. 
Although several works~\cite{liang2021swinir,sharif2020learning,kascenas2022denoising,rajesh2022evolutionary,khader2022medical} show less tendency toward overfitting and being data-driven, their performance is still highly dependent on the training samples. Since clean biomedical images for training are not available or available only in very limited quantities, this aspect of the current DNN techniques limits their generalizability to different sorts of biomedical images. 
\section{Proposed Method}\label{sec:proposed_preafce}
The core concept behind ADL is that image representations should exhibit resilience against noise and distortion. ADL comprises a denoiser and a discriminator, each employing a similar architecture referred to as \textit{Efficient-UNet}\footnote{
There are slight differences in the architecture of the denoiser's Efficient-UNet and the discriminator's, detailed in Section \ref{sec:Efficient}.}.   
ADL is trained by minimizing competing multiscale objectives between the denoiser and the discriminator. 
During training, we enforce the networks on keeping the edges, histogram, and pyramidal information of the reference/ground-truth data. 
The proposed loss functions are detailed in Section~\ref{sec:loss}.

\begin{figure*} 
    \centering
    \includegraphics[width=1\textwidth]{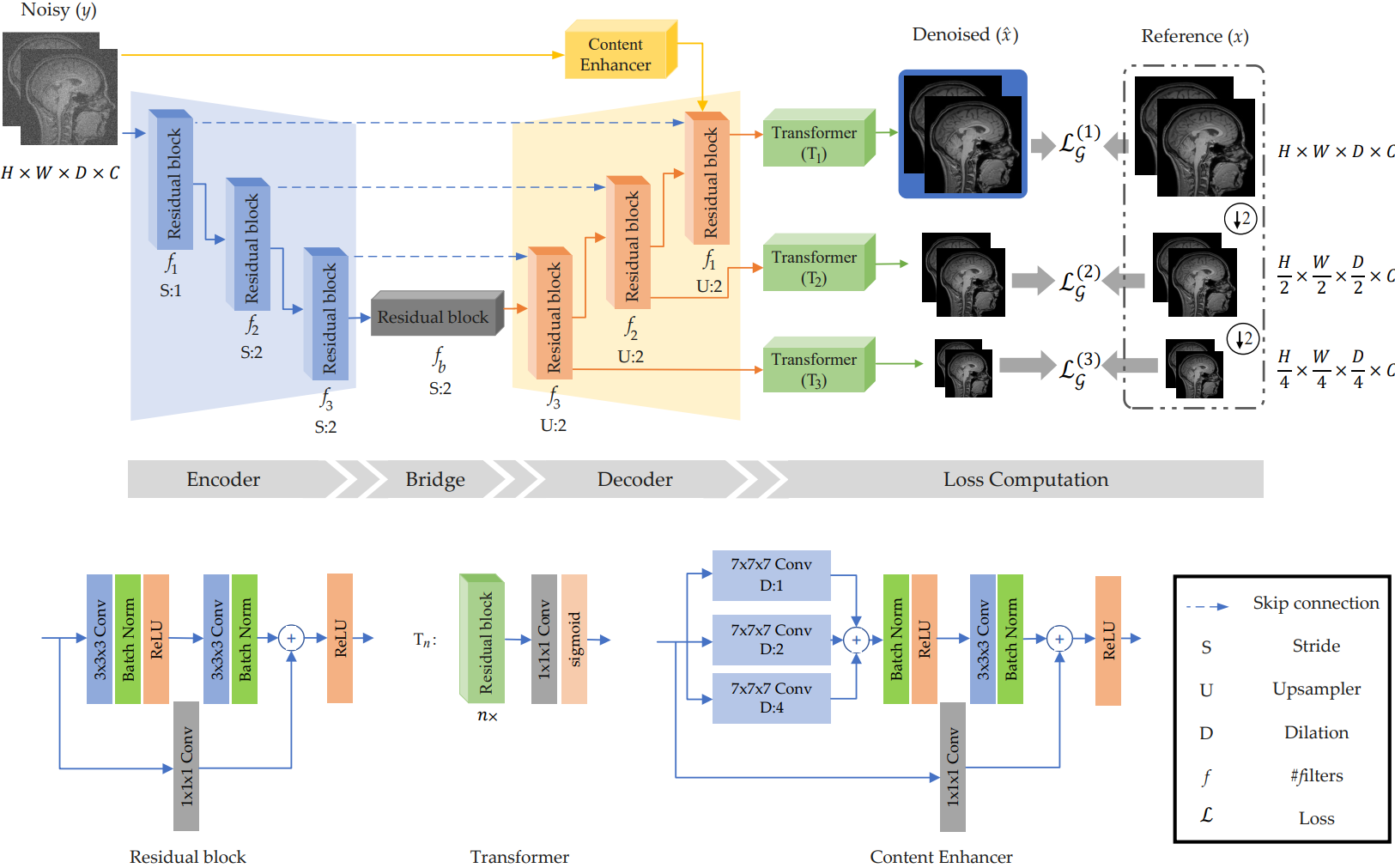}
\caption{Framework of the Efficient-UNet for denoiser with the training steps. 
Efficient-UNet is composed of the encoder, decoder, and Content Enhancer blocks. The output of the decoder at every scale is mapped into the image domain by a Transformer block. We then enforce consistency between the outputs of the decoder and their counterparts $x$. Low-level features further contribute to the denoised image by the Content Enhancer block. When the noise level is low, the filters of this block are activated, improving the convergence with no need for high-level features.
}
\label{fig:framework}
\end{figure*}

\subsection{Efficient-UNet}\label{sec:Efficient}
\begin{figure*} 
    \centering
    \includegraphics[width=0.9\textwidth]{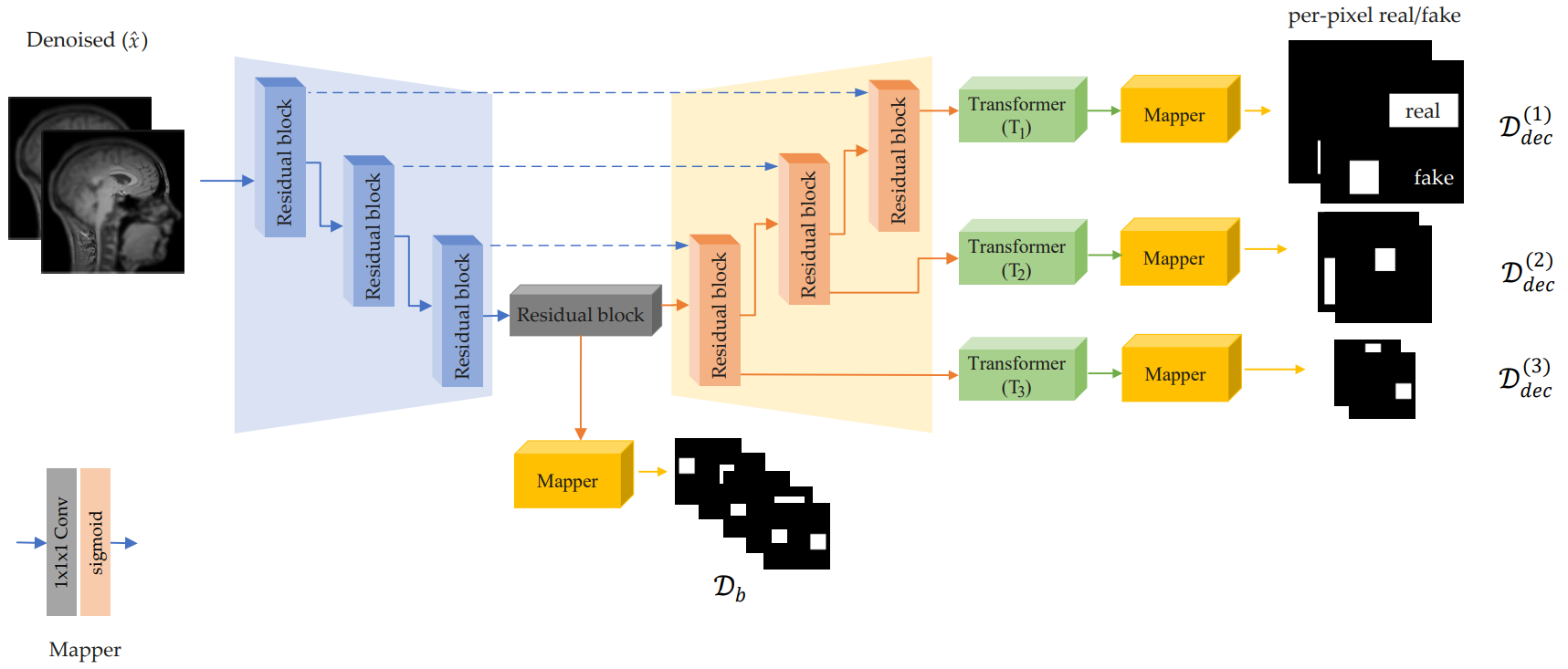}
\caption{Framework of the proposed Efficient-UNet for discriminator with the training steps.}
\label{fig:frameworkdisc}
\end{figure*}

Given a noise-free 3D (or 2D) visual data with $C$ channels represented as $\mathbf{x}\in\mathbb R^{H\times W\times D\times C}$ (or $\mathbf{x}\in\mathbb R^{H\times W\times C}$ for 2D). 
Let $\mathbf{y}\in\mathbb R^{H\times W\times D\times C}$ represent the observed visual data, which is generated by introducing white Gaussian noise (WGN) with a standard deviation, denoted $\sigma$, to the noise-free latent image $\mathbf{x}$.
$\mathbf{x}$ serves as a reference throughout the training process, and the noise level ($\sigma$) remains unspecified. 
The objective is to reconstruct $\mathbf{y}$ from $\mathbf{x}$. 
The architecture employed for training the denoiser is illustrated in Figure~\ref{fig:framework}, which relies on the Efficient-UNet framework. Efficient-UNet is composed of several components, including an encoder, a bridge, a decoder, a content enhancer, and transformers. Likewise, the architecture of Efficient-UNet for the discriminator is depicted in Figure~\ref{fig:frameworkdisc}. 
The \textit{encoder} unit is comprised of three consecutive residual blocks with strides one and two (denoted by $S$ in Figure~\ref{fig:framework}) that extract the features from the observed data. 
The number of filters in the encoder layers is doubled after each downsampling operation (i.e., $S=2$). Several studies~\cite{he2016deep,zhang2017beyond,sharif2020learning} have shown the effectiveness of residual blocks in deep learning. 
We use the residual blocks in our auto-encoder structure for denoiser prior modeling. To have large activation maps, we 
delay
the downsampling of the first layer, whose stride is 1. 
The extracted features by the encoder are then passed through a bridge with $f_{b}$ filters followed by the decoder unit that estimates the denoised input data $\mathbf{\hat x}$ in the feature domain. The features of each decoder layer are mapped into the image domain by the \textit{transformer} unit $T_{n}$ where $n$ is the index of the corresponding layer. 
The transformer unit is a set of $n$ residual blocks followed by a $1\times1\times1$ convolution and a sigmoid activation layer. The number of filters in the transformer layers is halved $n$ times\footnote{For example, the filter size of the residual block in $T_{1}$ is $\frac{f_{1}}{2}$. Similarly, we have $\{\frac{f_{3}}{2},\frac{f_{3}}{4},\frac{f_{3}}{8}\}$ for the residual blocks in $T_{3}$.} and the filter size of $1\times1\times1$ convolution layer is equal to $C$, which is the number of input channels. 
Throughout this study, the kernel size and dilation of the convolutional layers are 3 and 1, respectively, unless otherwise noted. 

Thus far, the only difference between the Efficient-UNet of the denoiser and that of the discriminator is the filter size of convolution layers, i.e., $\{f_{1}, f_{2}, f_{3}, f_{b}\}$. 
It is set as $\{96, 128, 160, 192\}$ for the denoiser and $\{48, 64, 80, 96\}$ for the discriminator.
The discriminator network needs a lower number of learning parameters compared to the denoiser. 
This is primarily because the discriminator functions as a classifier, responsible for computing the loss by distinguishing between noise-free and noisy, thereby determining if an instance is classified as noise-free or noisy. 
Since the discriminator is a classifier, the output of each transformer is mapped into a binary mask by a \textit{mapper} shown in Figure~\ref{fig:frameworkdisc}. 
The mapping layer comprises convolution layers with a filter size of 1 followed by a sigmoid activation layer to map the input to a binary mask. 
The denoiser also includes an additional component, a \textit{content enhancer}. This specific unit is strategically designed to retain sparse information, such as edges and textures, which may be minimally affected or remain unaffected by noise. 
As shown in Figure~\ref{fig:framework}, the content enhancer is a residual block, wherein the first layer extracts features by three different dilation values. 
Several studies~\cite{9292438,starck2007undecimated} have shown that the structural information, which is mainly edges and textures, exists at different scales. 
So we consider three convolutions of strides 1, 2, and 4 to extract such features from the input noisy data directly. 
In the absence of noise, this unit has more activated layers. 
The content enhancer layer is concatenated by the last layer of the decoder before feeding $T_{1}$. 
The denoised data $\mathbf{\hat x}$ is the output of layer $T_{1}$, shown by a blue rectangle in Figure~\ref{fig:framework}. 
It is also worth mentioning that the bias parameter in the proposed Efficient-UNet is deactivated since bias-free networks increase the linearity property by canceling the bias parameter\footnote{In~\cite{mohan2019robust}, it is shown that for any input $p$ and any non-negative constant $\alpha$, we have $h(\alpha p)$ is equal to $\alpha h(p)$ if the feed-forward network uses ReLU as the activation function with no additive constant terms (bias) in any layer.}~\cite{mohan2019robust}. 
Moreover, when the magnitude of the bias is much larger than that of filters, the model's generalizability is limited, so less prone to overfitting~\cite{zhang2021plug}.

\subsection{Multiscale loss framework}
\label{sec:loss}
ADL consists of two auto-encoders: a denoiser $\mathcal{G}$ and a discriminator $\mathcal{D}$. 
In a general setting, the parameters of adversarial networks are optimized by minimizing a two-player game in an alternating manner~\cite{goodfellow2014generative}:
\begin{equation}
\begin{array}{l}
\mathcal{L_D}=-\mathbb{E}_{\mathbf{x}}\{\text{log}\; \mathcal{D}(\mathbf{x})\} - \mathbb{E}_{\mathbf{y}}\{\text{log} \left(1-\mathcal{D}(\mathcal{G}(\mathbf{y}))\right)\},\\
\mathcal{L_G}= -\mathbb{E}_{\mathbf{y}}\{\text{log}\; \mathcal{D}(\mathcal{G}(\mathbf{y}))\}.
\end{array}
\label{eq:GD}
\end{equation}
Denoiser $\mathcal{G}$ aims to estimate a clean image while discriminator $\mathcal{D}$ distinguishes between reference $\mathbf{x}$ and denoised ${\mathcal{G}(\mathbf{y})}$ instances. 
Eq. \eqref{eq:GD} does not maintain textural, global, and local data representation. 
To this end, we propose novel multiscale loss functions for both the denoiser and the discriminator to replace the ones in Eq~\ref{eq:GD}. 
We minimize the loss of denoiser from coarse to fine resolution for enforcing the participation of the decoder's features in the resultant high-resolution denoised image. Likewise, we enforce the discriminator network to discriminate the target and the denoised images from bottom to top. 
Moving from coarse resolution towards fine one enhances structural features at the corresponding resolution, resulting in rich edge and textural information in the denoised data. 
\subsubsection{Denoiser's loss function ($\mathcal{L_G}$)}
We define the loss function of the denoiser as a combination of an ${\ell_1}$ loss (denoted by $\mathcal{L}_{\ell_1}$), a novel pyramidal textural loss $\mathcal{L}_{pyr}$, and a novel histogram loss $\mathcal{L}_{{H}ist}$ that are weighted by $\lambda_{\ell_1}$, $\lambda_{p}$, and $\lambda_{\mathcal{H}}$, respectively:

\begin{align}\label{eq:loss-denoiser}
\mathcal{L_G} & =  \lambda_{\ell_1} \mathcal{L}_{\ell_1} +\lambda_{p} \mathcal{L}_{pyr}+\lambda_{\mathcal{H}} \mathcal{L}_{\mathcal{H}ist}\\
& =-\mathbb{E}{\{\mid\mathcal{G}(\mathbf{y})-\mathbf{x}\mid}\}-\lambda_{p}\mathbb{E}\Big\{\sum_{j=1}^{J}\mid\Delta_{j}. \mathcal{G}(\mathbf{y}) - \Delta_{j}. \mathbf{x}\mid\Big\}
\\&
-\lambda_{\mathcal{H}}\mathbb{E}
\{\text{logcosh}\big(\mathcal{H}\left[\mathbf{\mathcal{G}(\mathbf{y})}\right] - \mathcal{H}\left[\mathbf{x}\right]\big)\}.
\end{align}

\begin{figure}
    \centering
    \includegraphics[width=0.5\textwidth]{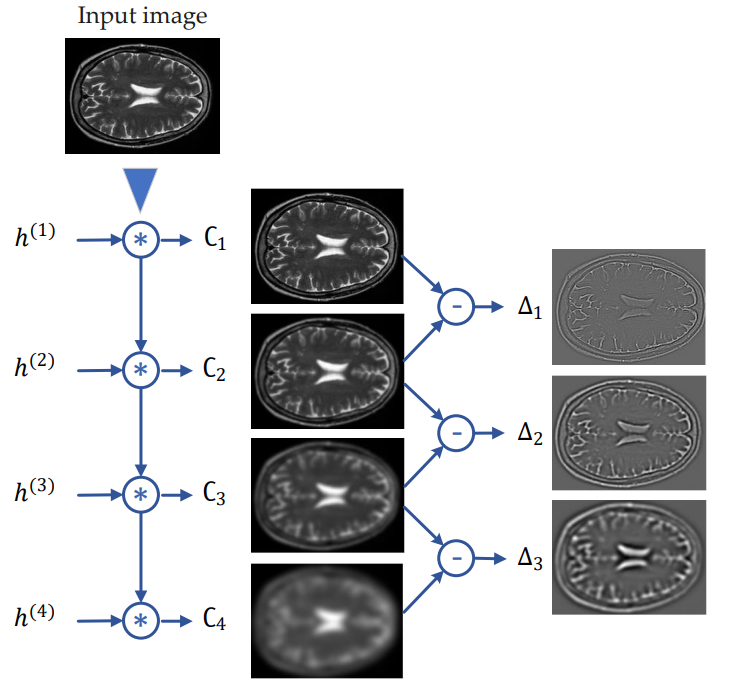}
\caption{ATW decomposition for image data. Since the 2D kernel $h$ is separable, the ATW decomposition for 3D image data is obtained by convolving three 1D cubic kernels in the x, y, and z directions. ATW is comprised of inherently low-pass filters that attenuate the side effects of noise.}
\label{fig:atw}
\end{figure}

\begin{itemize}
    \item[$-$] $\mathbf{\mathcal{L}_{\ell_1}}$: The data fidelity is an $\ell_1$-norm between the denoised and reference instances. Compared to the $\ell_2$-norm, the $\ell_1$-norm is more robust against outliers.
    \item[$-$] $\mathbf{\mathcal{L}_{pyr}}$: The pyramidal textural loss aims at preserving edges and textures in to-be-denoised images. 
    Unlike the conventional edge-preserving regularization terms 
    such as TV and its variants~\cite{papafitsoros2014combined,wang2019structural} that compute the differential of given noisy images, our goal is to measure the textural difference between to-be-denoised images and their corresponding reference only. 
    Traditional TVs use first-, second-, or higher-order-based derivative operators to filter noisy images. 
    The main drawback of such operators is to boost the pixels with high-level noise so that it may bias the loss towards dominant values. 
    For this reason, the images denoised by TV-like regularisation terms are often accompanied by smoothness in both textural and non-textural regions. 
    To cope with this problem, we introduce a pyramidal loss function using ATW~\cite{starck2007undecimated}. 
    ATW is a stationary Wavelet transform that decomposes an image into several levels by a cubic spline filter and then subtracts any two successive layers to obtain fine images with edges and textures. 
    The low-pass filters in ATW alleviate the side effects of noise, enabling us to include texture information of input images in the loss function. 
    Figure~\ref{fig:atw} briefs ATW. 
    In short, it decomposes an input image into $J$ levels by a cubic spline finite impulse response denoted by $h^{(1)}$. 
    Unlike the non-stationary multiscale transforms that downscale the images and then apply a filter, ATW upscales the kernel by inserting `$2^{j-1}-1$' zeros between each pair of adjacent elements in $h^{(1)}$, where $j$ denotes the $j$-th decomposition level. 
    Fine images with texture and edges are derived via subtraction of any two successive filtered images. Further details can be found in~\cite{starck2007undecimated,9292438}. 
    In Eq.~\ref{eq:loss-denoiser}`$\Delta_{j}.$' denotes the textural image at the $j$-th level derived by ATW. $J$ is a positive integer that denotes the number of decomposition levels. Typically, four decomposition levels ($J=4$) have been able to extract the majority of edges and textures laid in a wide range of sigma values~\cite{9292438}. 
    \item[$-$] $\mathbf{\mathcal{L}_{\mathcal{H}ist}}$: The histogram loss assures that the histograms of $\mathcal{G}(\mathbf{y})$ and $\mathbf{x}$ are close to each other. This term maintains the global structure of $\mathcal{G}(\mathbf{y})$ with respect to $\mathbf{x}$ since the added edges and texture information (by ATW) may change the overall histogram of the denoised instance. 
    To compute this loss, we first compute the histogram of both $\mathcal{G}(\mathbf{y})$ and $\mathbf{x}$ denoted by $\mathcal{H}$. 
    We then use a strictly increasing function that computes the loss between $\mathcal{H}\left[\mathbf{\mathcal{G}(\mathbf{y})}\right] $ and $\mathcal{H}\left[\mathbf{x}\right]$. To this end, we use $\text{logcosh}(p)=\text{log(cosh(}p\text{))}$ that is approximately equal to $\frac{p^2}{2}$ for small $p$ and to $\text{log}(p)-\text{log}(2)$ for large $p$\footnote{If the number of color channels is more than one, i.e. $C>1$, we first compute the histogram loss between corresponding channels and then take the average of losses as the final histogram loss.}. 
    Note that we use the histogram loss for preserving the global structure between the denoised image and its reference, and this is different from other histogram-based works like~\cite{zuo2013texture,talebi2021projected}. In \cite{zuo2013texture}, a gradient histogram is used for texture preservation, and Delbracio~\etal~\cite{talebi2021projected} proposed a DNN-based histogram as a fidelity term.
\end{itemize}

In Efficient-UNet, features at fine-resolution maps are reconstructed from their coarser ones (Figure~\ref{fig:framework}). Thus, an inefficiency in coarser spatial scales (or, equivalently, higher-level feature maps) may yield poor resultant feature maps. To cope with this problem, we propose a pyramidal version of the loss function defined in Eq.~\ref{eq:loss-denoiser} that maintains
the consistency between $\mathcal{G}(\mathbf{y})$ and its counterpart $\mathbf{x}$ at each spatial scale. 
To compute the proposed multiscale loss, we generalize the transformer unit, introduced in Section~\ref{sec:Efficient}, to lower spatial scales (here, two scales $T_2$ and $T_3$) and then get the corresponding denoised image data. 
We compute the loss function presented in Eq.~\ref{eq:loss-denoiser} at every scale and finally take an average of the losses to yield the denoiser loss:
\begin{equation}
\mathcal{L_G}= \sum_{s=1}^{3}\left(\lambda_{\ell_1} \mathcal{L}^{(s)}_{\ell_{1}} +\lambda_{e} \mathcal{L}^{(s)}_{edge}+\lambda_{\mathcal{H}} \mathcal{L}^{(s)}_{\mathcal{H}ist}\right),
\label{eq:loss-denoisermulti}
\end{equation} 
where superscript $s$ denotes the scale of the data pyramid. The total number of scales is set to 3 in this study. 

\subsubsection{Discriminator's loss function ($\mathcal{L_D}$)} Most discriminators are based on an encoder that determines whether the input instance is fake or real. 
Schonfeld~\etal \cite{schonfeld2020u} have shown that using an U-Net structure for a discriminator could increase the generator's performance (here, the denoiser). Borrowed a similar concept from~\cite{schonfeld2020u}, we propose the following loss function for the discriminator (see Figure~\ref{fig:frameworkdisc}):
\begin{equation}
\mathcal{L_D}= \mathcal{L}_{\mathcal{D}_b}+\sum_{s=1}^{3}\mathcal{L}^{(s)}_{\mathcal{D}_{dec}},
\label{eq:loss-discmulti}
\end{equation}
where $\mathcal{L}_{\mathcal{D}_b}$ is the bridge loss 
defined as 
\begin{equation}
\mathcal{L}_{\mathcal{D}_b}=-\mathbb{E}\Big\{\sum_{m,n,d}\text{min}(0,-1+{\mathcal{D}_b}(\mathbf{x})\big|_{m,n,d}\Big\}-\mathbb{E}\Big\{\sum_{m,n,d}\text{min}(0,-1-{\mathcal{D}_b}(\mathbf{\hat{x}})\big|_{m,n,d}\Big\}.
\label{eq:loss-discmulti1}
\end{equation}
Here, subscripts $\{m,n,d\}$ denote discriminator decision at pixel $(m,n,d)$. Likewise, $\mathcal{L}_{\mathcal{D}_{dec}}$ in Eq.~\ref{eq:loss-discmulti} is defined as follows:
\begin{equation}
\mathcal{L}^{(s)}_{\mathcal{D}_{dec}}=-\mathbb{E}\Big\{\sum_{m,n,d}\text{min}(0,-1+{\mathcal{D}_{dec}}(\mathbf{x})\big|_{m,n,d}\Big\}-\mathbb{E}\Big\{\sum_{m,n,d}\text{min}(0,-1-{\mathcal{D}_{dec}}(\mathbf{\hat{x}})\big|_{m,n,d}\Big\},\:s\in\{1,2,3\}.
\label{eq:loss-discmulti2}
\end{equation}
\subsection{Training details}\label{sec:implementation}
We implemented ADL for both 2D and 3D image data. To show the generalizability of the denoiser, we trained the 2D model on ImageNet~\cite{russakovsky2015imagenet}, whose data distribution is different from that of biomedical images. 
The training and the validation sets contain 1,281,168 and 50,000 natural images of size $224\times224$, respectively. 
The denoiser model was trained on color and gray-scale images separately. 
We augmented the data by rotation, $\theta\in\{90^{\circ}, 180^{\circ}, 270^{\circ}\}$, and flips. 
We added zero-mean WGN to input images with $\sigma$ randomly selected in the interval [0,55]. 
The noisy images were not clipped into the interval [0,255], ensuring that the added noise distribution was still WGN. 
The 3D model was trained on the IXI Dataset\footnote{\url{https://brain-development.org/ixi-dataset}}, containing approximately 570 8-bit T1, T2, and PD-weighted MR images acquired at three different hospitals in London: Hammersmith Hospital using a Philips 3T system, Guy’s Hospital using a Philips 1.5T system, and Institute of Psychiatry using a GE 1.5T system. 
The size of MRI volumes is 256$\times$256$\times n$, where $n\in[28,150]$.
We randomly cropped the volumes into sub-volumes of 128$\times$128$\times$48 to facilitate the training. 
we added zero-mean white Gaussian ($\sigma\in[0,55]$), Rician ($\sigma\in[0,25]$ and $\nu\in[0,15]$), and Rayleigh ($\sigma\in[0,25]$) noise to input MRI volumes. 
Likewise, in the 2D training, the values of the parameters in the above-mentioned intervals were randomly selected. 
We used the Adam algorithm~\cite{kingma2014adam} as an optimizer with the \textit{ReduceLRonPlateau} scheduler\footnote{\url{https://www.tensorflow.org/api_docs/python/tf/keras/callbacks/ReduceLROnPlateau}}. 
The optimization started with a learning rate of $10^{-4}$ and was reduced (by the ReduceLRonPlateau scheduler) until the PSNR of validation stopped improving. 
The weights in Eq.~\ref{eq:loss-denoisermulti}, i.e. $\{\lambda_{\ell_1}$, $\lambda_{e}$, $\lambda_{\mathcal{H}}\}$, were set to 1.

\textbf{Reproducibility}: The reference implementation of ADL is based on TensorFlow. All the experiments were conducted on 4 A-100 GPUs. 
The implementation and pre-trained models are available at \url{https://github.com/mogvision/ADL}. 
We also provided PyTorch and Colab implementation of ADL on GitHub.

\renewcommand{\tabcolsep}{2.5pt}
\begin{table*}
\caption{PSNR (dB) results (average $\pm$ standard deviation) of different methods on the image-based datasets for noise levels 10, 15, 25, and 35. The best and second best results are highlighted in \textbf{boldface} and in \textit{italic}, respectively.} 
    \label{table:psnr}
\centering 
\begin{tabular}{lccccccccc} %
\toprule
   Dataset& Noise Level & BM3D~\cite{makinen2020collaborative} & DRAN~\cite{sharif2020learning} & DnCNN-S~\cite{zhang2017beyond} & SwinIR~\cite{liang2021swinir} & Proposed ADL\\
\toprule
& 10 & 37.4$\pm$1.3 & 37.5$\pm$1.1 & 38$\pm$1.4 & \textit{38.4$\pm$1.4} & \textbf{38.8$\pm$1.3}\\
& 15 & 35.9$\pm$1.3 & 36.9$\pm$1.4 & 35.8$\pm$1.7 & \textit{37.1$\pm$1.4} & \textbf{37.5$\pm$1.3}\\
\verb//HAM10000~\cite{DBW86T_2018}& 25 & 34.4$\pm$1.5 & 34.5$\pm$1.5 & 34.9$\pm$1.6 & \textit{35.5$\pm$1.6} & \textbf{36.1$\pm$1.4}\\
& 35 & 33.4$\pm$1.6 & 32.6$\pm$1.5 & 33.9$\pm$1.7 & \textit{34.8$\pm$1.7} & \textbf{35.2$\pm$1.5}\\

\midrule
& 10 & 38.6$\pm$ 1.1 & 38.6$\pm$1.6 & 38.8$\pm$1.1 & \textit{38.9$\pm$1.2} & \textbf{39.6$\pm$1.1}\\
& 15 & 37.6$\pm$1.1 & 37.9$\pm$1.8 & 38.5$\pm$1.2 & \textbf{38.7$\pm$1.2} & \textit{38.6$\pm$1.2}\\
\verb//Chest X-Ray~\cite{kermany2018identifying}& 25 & 35.4$\pm$0.9 & 36.5$\pm$1.7 & 35.8$\pm$1 & \textit{36.9$\pm$1.1} & \textbf{37.1$\pm$1.2}\\
& 35 & 33.8$\pm$0.9 & 35$\pm$1.8 &34.5$\pm$0.9 & \textit{35.1$\pm$1} & \textbf{35.7$\pm$1.1}\\
\midrule
& 10 & 30.5$\pm$0.3 & 31.2$\pm$0.5 & 31.1$\pm$0.3 & \textit{31$\pm$0.3} & \textbf{31.5$\pm$0.3} \\
& 15 & 28.2$\pm$0.3 & 29$\pm$0.5 & 28.9$\pm$0.3 & \textit{29.1$\pm$0.3} & \textbf{29.3$\pm$0.3}\\
\verb//EM~\cite{isbi2021}& 25 & 25.6$\pm$0.4 & 26.5$\pm$0.6 & 26.3$\pm$0.4 & \textit{26.6$\pm$0.4} & \textbf{27$\pm$0.4}\\
& 35 & 24$\pm$0.4 & 24.9$\pm$0.5 & 24.7$\pm$0.4& 2\textit{5$\pm$0.5} & \textbf{25.3$\pm$0.4}\\
\bottomrule
\end{tabular}
\end{table*}

\renewcommand{\tabcolsep}{2.5pt}
\begin{table*}
\caption{SSIM results (average $\pm$ standard deviation) of different methods on the image-based datasets for noise levels 10, 15, 25, and 35.} 
    \label{table:ssim}
\centering 
\begin{tabular}{lccccccccc} %
\toprule
   Dataset& Noise Level & BM3D~\cite{makinen2020collaborative} & DRAN~\cite{sharif2020learning} & DnCNN-S~\cite{zhang2017beyond} & SwinIR~\cite{liang2021swinir} & Proposed ADL\\
\toprule
& 10 & 0.91$\pm$0.02 & \textit{0.93$\pm$0.02} & 0.91$\pm$0.02 & 0.92$\pm$0.02 & \textbf{0.95$\pm$0.02}\\
& 15 & 0.90$\pm$0.03 & 0.91$\pm$0.03 & 0.91$\pm$0.03 & \textit{0.92$\pm$0.02} & \textbf{0.94$\pm$0.02}\\
\verb//HAM10000~\cite{DBW86T_2018}& 25 & 0.88$\pm$0.03 & 0.85$\pm$0.04 & 0.86$\pm$0.04 & \textit{0.90$\pm$0.03} & \textbf{0.93$\pm$0.02}\\
& 35 & 0.85$\pm$0.04 & 0.82$\pm$0.05 & 0.84$\pm$0.05 & \textit{0.86$\pm$0.05} & \textbf{0.93$\pm$0.02}\\

\midrule
& 10 & 0.94$\pm$0.02 & \textit{0.95$\pm$0.02} & 0.94$\pm$0.02 & \textbf{0.96$\pm$0.02} & \textbf{0.96$\pm$0.02}\\
& 15 & 0.93$\pm$0.02 & 0.94$\pm$0.03 & 0.93$\pm$0.02 & \textbf{0.94$\pm$0.02} & \textbf{0.94$\pm$0.02}\\
\verb//Chest X-Ray~\cite{kermany2018identifying}& 25 & 0.91$\pm$0.02 & 0.92$\pm$0.04 & 0.92$\pm$0.02 & 0.92$\pm$0.02 & \textbf{0.93$\pm$0.02}\\
& 35 & 0.89$\pm$0.02 & 0.90$\pm$0.04 & 0.90$\pm$0.02 & \textit{0.91$\pm$0.03} & \textbf{0.92$\pm$0.03}\\
\midrule
& 10 & 0.91$\pm$0.02& 0.91$\pm$0.03 & 0.89$\pm$0.01 &  0.91$\pm$0.03 & \textbf{0.93$\pm$0.02}\\
& 15 & 0.85$\pm$0.02 & \textit{0.87$\pm$0.02} & 0.86$\pm$0.01 & \textit{0.87$\pm$0.02} & \textbf{0.88$\pm$0.01}\\
\verb//EM~\cite{isbi2021}& 25 & 0.76$\pm$0.03 & \textit{0.79$\pm$0.04} & 0.78$\pm$0.03 & \textit{0.79$\pm$0.02} & \textbf{0.80$\pm$0.02}\\
& 35 & 0.69$\pm$0.04 & 0.74$\pm$0.04 & 0.72$\pm$0.03 & \textit{0.73$\pm$0.03} & \textbf{0.75$\pm$0.03}\\
\bottomrule
\end{tabular}
\end{table*}

\section{Experimental Results}\label{sec:experimental}
\begin{figure}
    \centering
    \includegraphics[width=0.45\textwidth]{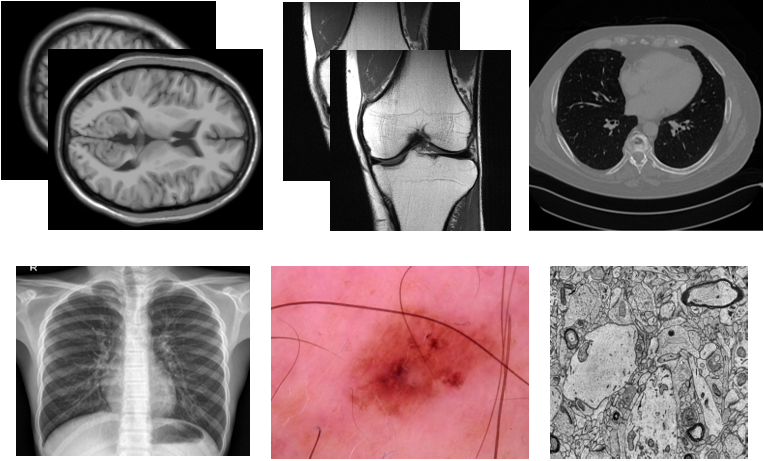}
\caption{A sample of each dataset used for evaluation in this study. From top to bottom and left to right: 3D Brain MRI~\cite{cocosco1997brainweb}, 3D knee MRI~\cite{zbontar2018fastmri},  chest X-ray~\cite{kermany2018identifying}, dermatoscopic RGB~\cite{DBW86T_2018}, and EM~\cite{isbi2021}.}
\label{fig:datasets}
\end{figure}

\textbf{Datasets}: We used five biomedical imaging datasets from different modalities (3D brain and knee MRI), EM, X-ray, and dermatoscopy to evaluate our model. The datasets are summarized below and an example image from each dataset is shown in Figure~\ref{fig:datasets}. Note that our method was trained on data (ImageNet for 2D and IXI for 3D) fully independent of these five test datasets. The training set was used for training the competing deep learning techniques.

\begin{itemize}
    \item \textbf{BrainWeb} database (3D)~\cite{cocosco1997brainweb} contains simulated 3D brain MRI based on healthy and multiple sclerosis (MS) anatomical models. MRI volumes have been simulated modelling three pulse sequences (T1-, T2- and PD-weighted), five slice thicknesses (1, 3, 5, 7, and 9 mm; pixel size is 1 mm$^2$), clean and noisy samples with five levels of Rician noise (1\%, 3\%, 5\%, 7\%, 9\%), and three levels of intensity non-uniformity (INU) 0\%, 20\%, and 40\%. In total, this results in 30 reference (clean, no INU) and 510 noisy MRI volumes of size 181$\times$217$\times n$, where $n = 181, 60, 36,26, 20$ depending on slice thickness. 
    \item The NYU \textbf{fastMRI} Initiative database (fastmri.med.nyu.edu) (3D) ~\cite{zbontar2018fastmri} contains PD-weighted knee MRI scans with and without fat suppression with in-plane size $320\times320$ and the number of slices varying from 27 to 45. We randomly sampled 200 volumes from the database for evaluation.
    \item \textbf{MitoEM Challenge} ~\cite{isbi2021} contains multibeam scanning EM volume taken from Layer II in temporal lobe of adult human. The volume was 1000 $\times$ 4096 $\times$ 4096  with the voxel size of 30 nm $\times$ 8 nm $\times$ 8 nm. Due to large slice separation, we considered slices as 2D images, resized them to 2048 $\times$ 2048 to reduce noise and sampled $512\times512$ non-overlapping patches - 4,008 for training and 646 for test.
    \item \textbf{Chest X-ray} database ~\cite{kermany2018identifying} contains routine clinical images (anterior-posterior, $600\times900$) from pediatric patients (healthy, pneumonia) of one to five years old from Guangzhou Women and Children’s Medical Center, Guangzhou. The dataset was split into 5,232 training and 624 test images. 
    \item \textbf{The HAM10000 }~\cite{DBW86T_2018} database contains 10,015 dermatoscopic RGB images ($400\times600\times3$) of pigmented skin lesions from different populations, acquired and stored by different modalities. We set the training/test set ratio to 80\%-20\%.
\end{itemize}

\begin{figure*}
    \centering
    \subfigure[Reference]{
        \centering
        \includegraphics[width=0.23\textwidth, height=0.35\textwidth]{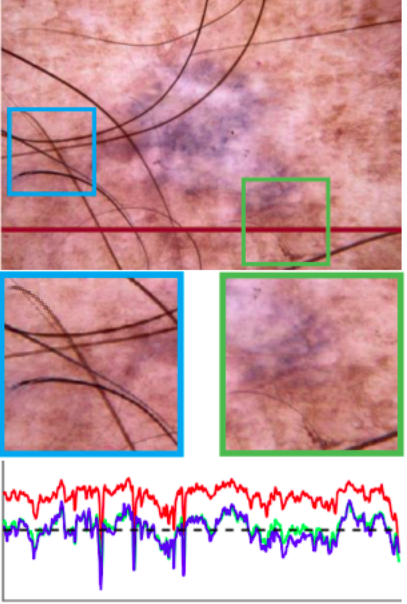}
        \label{fig:2dskin_ref}
    }
    \subfigure[Noisy]{
        \centering
        \includegraphics[width=0.23\textwidth, height=0.35\textwidth]{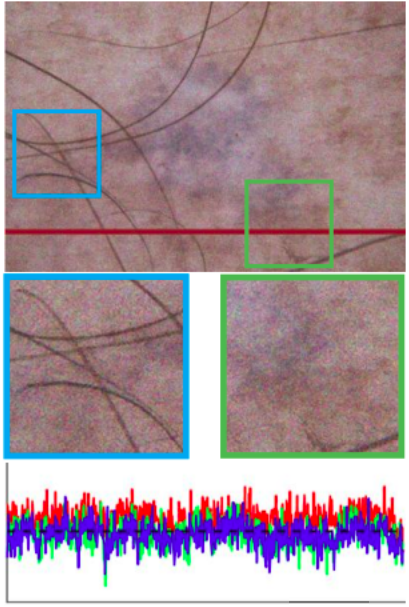}
        \label{fig:2dskin_noisy}
    }
    \subfigure[BM3D]{
        \centering
        \includegraphics[width=0.23\textwidth, height=0.35\textwidth]{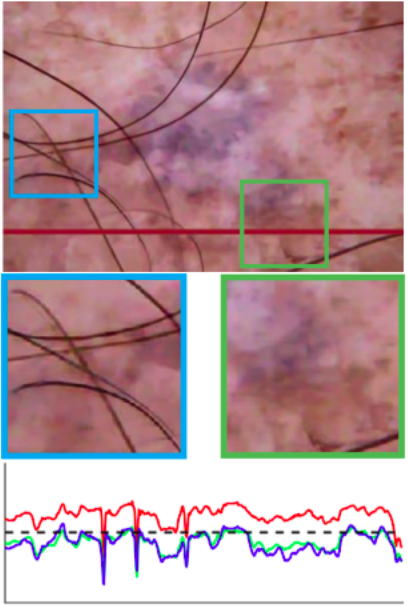}
        \label{fig:2dskin_bm3d}
    }\\
    \subfigure[SwinIR]{
        \centering
        \includegraphics[width=0.23\textwidth, height=0.35\textwidth]{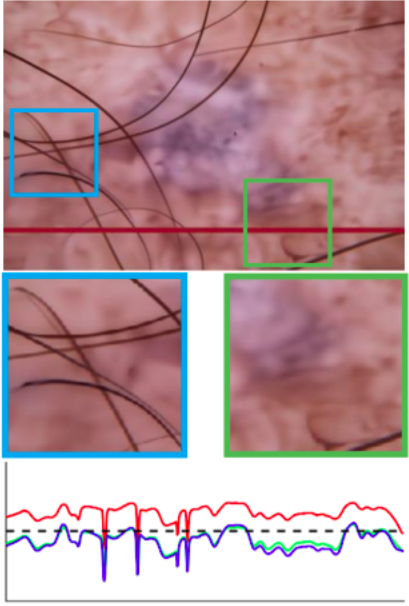}
        \label{fig:2dskin_swin}
    }
    \subfigure[Proposed ADL]{
        \centering
        \includegraphics[width=0.23\textwidth, height=0.35\textwidth]{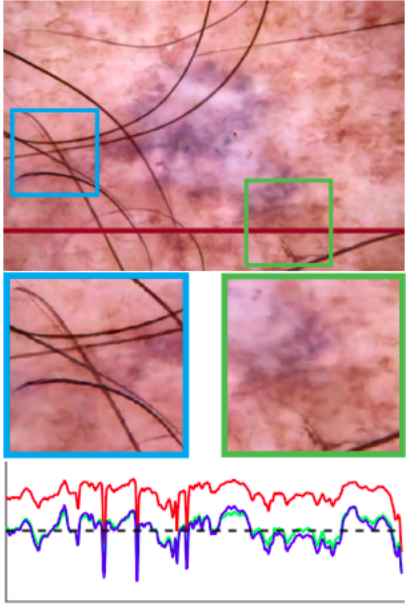}
        \label{fig:2dskin_adl}
    }
\caption{Color image denoising results
of different methods on  dermatoscopic RGB~\cite{DBW86T_2018}
}
\label{fig:visual2Dskin}
\end{figure*}

\begin{figure*}
    \centering
    \subfigure[Reference]{
        \centering
        \includegraphics[width=0.23\textwidth, height=0.35\textwidth]{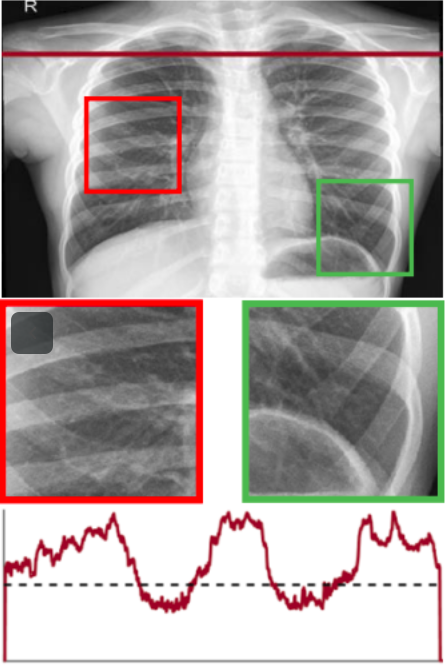}
        \label{fig:2dchest_ref}
    }
    \subfigure[Noisy]{
        \centering
        \includegraphics[width=0.23\textwidth, height=0.35\textwidth]{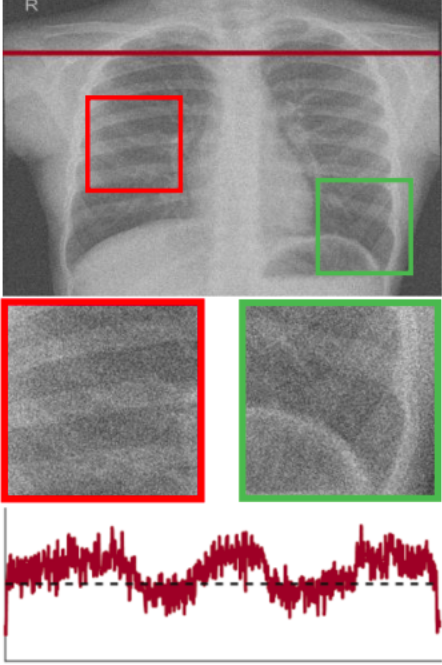}
        \label{fig:2dchest_noisy}
    }
    \subfigure[BM3D]{
        \centering
        \includegraphics[width=0.23\textwidth, height=0.35\textwidth]{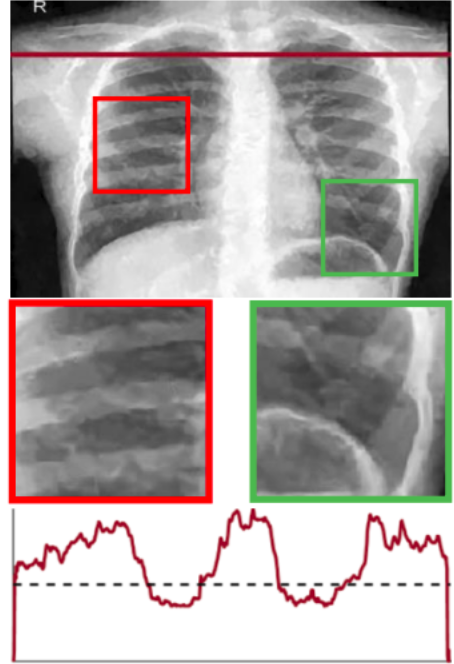}
        \label{fig:2dchest_bm3d}
    }\\
    \subfigure[SwinIR]{
        \centering
        \includegraphics[width=0.23\textwidth, height=0.35\textwidth]{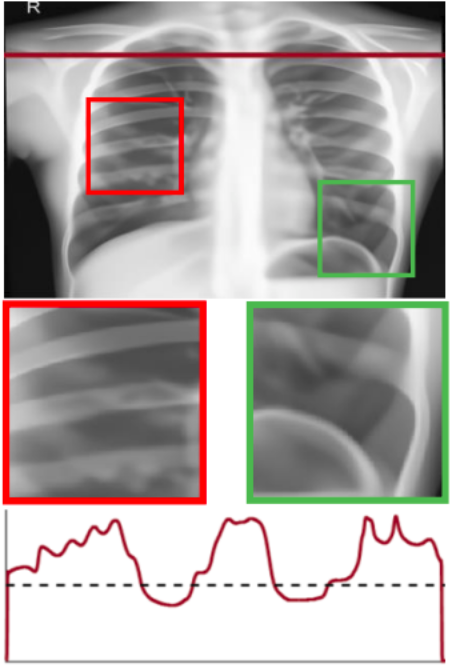}
        \label{fig:2dchest_swin}
    }
    \subfigure[Proposed ADL]{
        \centering
        \includegraphics[width=0.23\textwidth, height=0.35\textwidth]{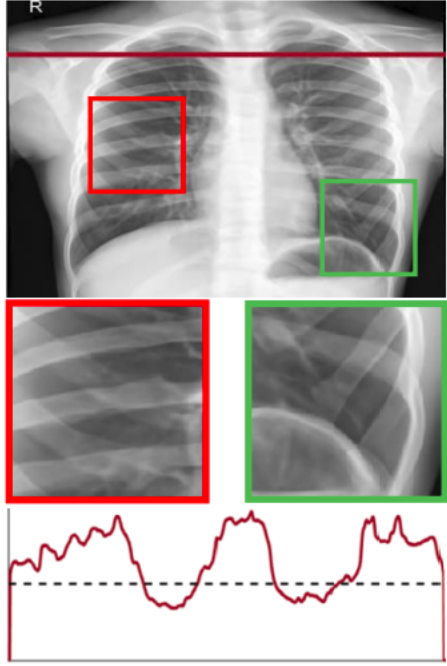}
        \label{fig:2dchest_adl}
    }
\caption{Gray-scale image denoising results
of different methods on chest X-ray~\cite{kermany2018identifying}
}
\label{fig:visual2Dchest}
\end{figure*}

\begin{figure*}
    \centering
    \subfigure[Reference]{
        \centering
        \includegraphics[width=0.23\textwidth, height=0.35\textwidth]{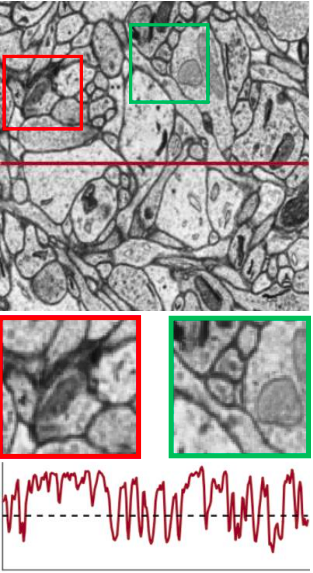}
        \label{fig:em_ref}
    }
    \subfigure[Noisy]{
        \centering
        \includegraphics[width=0.23\textwidth, height=0.35\textwidth]{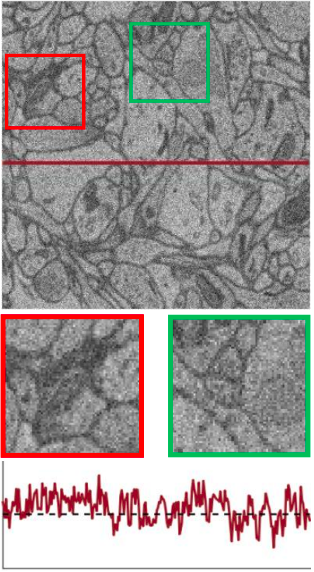}
        \label{fig:2dem_noisy}
    }
    \subfigure[BM3D]{
        \centering
        \includegraphics[width=0.23\textwidth, height=0.35\textwidth]{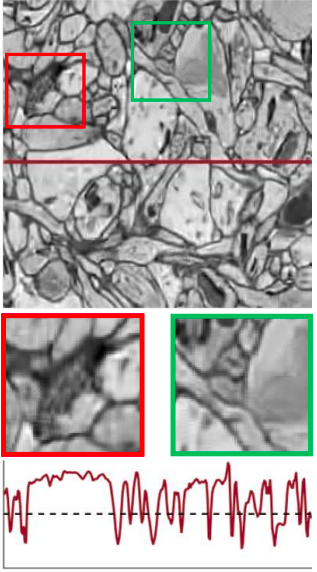}
        \label{fig:2dem_bm3d}
    }\\
    \subfigure[SwinIR]{
        \centering
        \includegraphics[width=0.23\textwidth, height=0.35\textwidth]{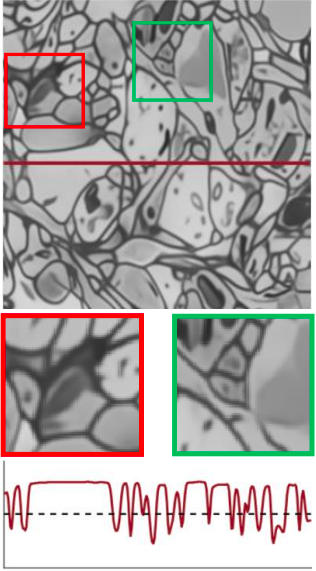}
        \label{fig:2dem_swin}
    }
    \subfigure[Proposed ADL]{
        \centering
        \includegraphics[width=0.23\textwidth, height=0.35\textwidth]{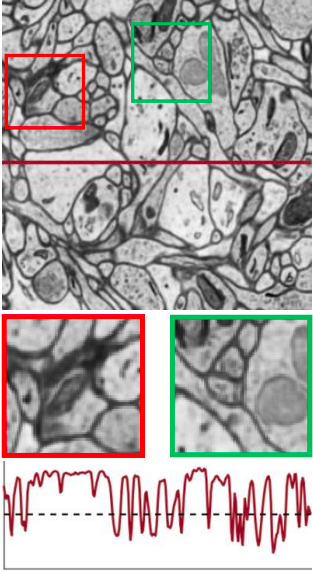}
        \label{fig:2dem_adl}
    }
\caption{Gray-scale image denoising results
of different methods on  dermatoscopic  EM~\cite{isbi2021}
}
\label{fig:visual2Dem}
\end{figure*}

For 2D databases, we compared the proposed ADL to improved BM3D~\cite{makinen2020collaborative}\footnote{\url{https://pypi.org/project/bm3d/}} as the conventional model-based method, and three deep learning-based methods, Dynamic Residual Attention Network (DRAN)~\cite{sharif2020learning}, DnCNN-S~\cite{zhang2017beyond}, and recently-developed SwinIR~\cite{liang2021swinir}. We compared the 3D version of our model to BM4D~\cite{maggioni2012nonlocal}, which is widely used for denoising 3D biomedical image data.


\subsection{Experiments with 2D data with synthetic noise}

\begin{table}
\renewcommand{\tabcolsep}{4.8pt} 
\caption{Average PSNR (dB) and SSIM results  of different methods  for low-dose CT denoising.} 
    \label{table:resultsCTLowDose}
\centering 
\begin{tabular}{lcccccccc} %
\toprule
Dataset & \#Images & Metric & Quarter Dose & RED-CNN & Q-AE & WGAN-VGG & CNCL & ADL\\
&&&&\cite{chen2017low}& & & \cite{geng2021content} &(ours)\\
\toprule
\multirow{2}{*}{B30 (1 mm$^2$)}&\multirow{2}{*}{1176}&PSNR$\uparrow$& 36.37 & 38.05 & 3 & 4 & 5 &39.83\\
&&SSIM$\uparrow$& 0.873 & 0.911 & 3 & 4 & 5 &0.948\\
\midrule
\multirow{2}{*}{D45 (1 mm$^2$)}&\multirow{2}{*}{1258}&PSNR$\uparrow$& 28.53& 30.82 & 3 & 4 & 5 & 33.66\\
&&SSIM$\uparrow$& 0.643 & 0.731 & 3 & 4 & 5 & 0.818\\
\midrule
\multirow{2}{*}{B30 (3 mm$^2$)}&\multirow{2}{*}{520}&PSNR$\uparrow$& 41.32 & 42.58 & 3 & 4 & 5 & 43.69\\
&&SSIM$\uparrow$& 0.951 & 0.968 & 3 & 4 & 5 & 0.978\\
\bottomrule
\end{tabular}
\end{table}

WGN with $\sigma\in [10,15,25,35]$ were added into the test images of the skin~\cite{DBW86T_2018}, chest~\cite{kermany2018identifying}, and EM~\cite{isbi2021} datasets. 
The PSNR (dB) and SSIM results for different methods are represented in Tables~\ref{table:psnr} and~\ref{table:ssim}, respectively.
According to the PSNR and SSIM results, one can see that the deep learning techniques yield better results compared to the model-based BM3D. 
The results of ADL are on par with those of the other methods, and it outperforms the other competing methods by a large margin for almost any noise level.
The standard deviation measures the compactness of the denoising results.
Although our method was not trained on the training sets (see Section~\ref{sec:experimental}), its PSNR's standard deviation results are on par with the conventional techniques. 
The SSIM results report that ADL gained the most compact results over all the noise levels. 
Figures~\ref{fig:visual2Dskin}--\ref{fig:visual2Dem} show denoising examples of different methods on the 2D datasets with noise level 35. To save space, we only show the results of model-based BM3D and SwinIR results, as SwinIR was the best among the competing deep learning-based techniques. 
The example images are accompanied by profile lines.
In the HAM10000 dataset (see Figure~\ref{fig:visual2Dskin}),  ADL could recover much sharper edges than BM3D and SwinIR while maintaining the contrast of the reference image. 
The RGB profile lines of ADL have the highest similarity with the reference ones, while the competing methods failed to preserve the contrast.

\begin{figure*}
    \subfigure[Reference]{
        \centering
        \includegraphics[width=0.23\textwidth, height=0.18\textwidth]{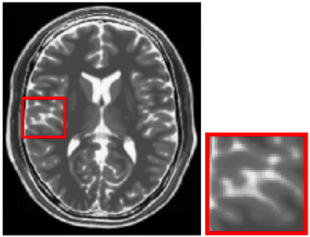}
        \label{fig:3d_ref}
    }
    \subfigure[Noisy]{
        \centering
        \includegraphics[width=0.23\textwidth, height=0.18\textwidth]{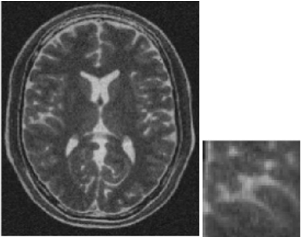}
        \label{fig:3d_noisy}
    }
    \subfigure[BM4D]{
        \centering
        \includegraphics[width=0.23\textwidth, height=0.18\textwidth]{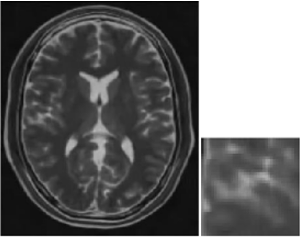}
        \label{fig:3d_bm3d}
    }
    \subfigure[Proposed ADL]{
        \centering
        \includegraphics[width=0.23\textwidth, height=0.18\textwidth]{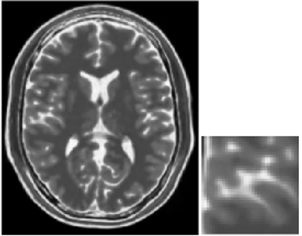}
        \label{fig:3d_adl}
    }
\caption{Denoising results of BM4D and the proposed ADL on the BrainWeb data with noise level 9 and RF 40\%}
\label{fig:visualcomp3d}
\end{figure*}

In the Chest X-Ray dataset (Figure~\ref{fig:visual2Dchest}), although the profile lines of all the methods were highly correlated to the reference one, ADL preserved more textural information compared to the others. 
BM3D resulted in blurred texture regions and SwinIR smoothed out the bones while the proposed ADL recovered more fine details and textures.
The same trend is seen in the EM results depicted in Figure~\ref{fig:visual2Dem}, where SwinIR yielded a toy-like result and BM3D blurred the borders. 
The smoothness of BM3D and SwinIR is apparent in their profile lines while the profile line of ADL tracked the reference one tightly. 
This means that ADL could successfully preserve the contrast and edges of the reference.

\begin{figure}
    \centering
    \includegraphics[width=0.9\textwidth]{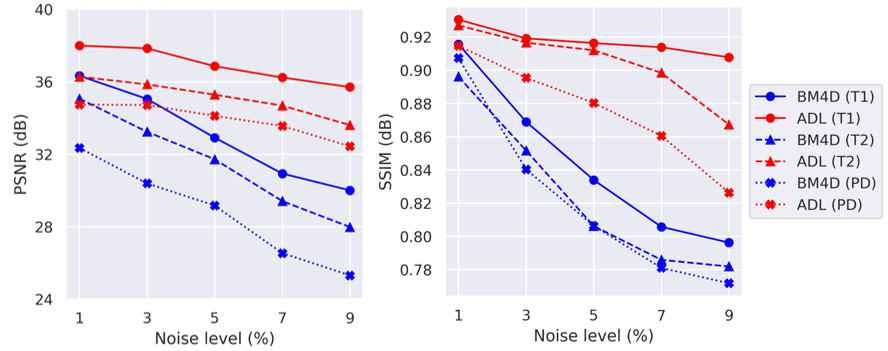}
\caption{Average PSNR(dB) and SSIM of BM4D~\cite{maggioni2012nonlocal} and the proposed ADL on the BrainWeb database~\cite{cocosco1997brainweb}.}
\label{fig:mri3d}
\end{figure} 

\begin{table}
\caption{PSNR (dB) and SSIM results (average $\pm$ standard deviation) of BM4D~\cite{maggioni2012nonlocal} and the proposed ADL on the 3D MRI knee dataset~\cite{zbontar2018fastmri} for noise levels 10, 15, 25, and 35. The best results are highlighted in \textbf{boldface}.} 
    \label{table:resultslnee}
\centering 
\begin{tabular}{lcccccccc} %
\toprule
    \multirow{3}{*}{
    \parbox[c]{.04\linewidth}{\centering Dataset}}&
    \multirow{3}{*}{
    \parbox[c]{.09\linewidth}{\centering Noise\\Level}}
    & \multicolumn{2}{c}{PSNR$\uparrow$} &
    &\multicolumn{2}{c}{SSIM$\uparrow$} \\ 
    \cmidrule{3-4} \cmidrule{6-7}
 && \multicolumn{1}{c}{BM4D} & \multicolumn{1}{c}{ADL} & & \multicolumn{1}{c}{BM4D} &  \multicolumn{1}{c}{ADL}\\
  && \multicolumn{1}{c}{\cite{maggioni2012nonlocal}} &  \multicolumn{1}{c}{(ours)} & & \multicolumn{1}{c}{\cite{maggioni2012nonlocal}} & \multicolumn{1}{c}{(ours)}\\
\toprule
& 10 & 37.1$\pm$1.8 &\textbf{38.3$\pm$1}  && 0.89$\pm$0.04  & \textbf{0.92$\pm$0.02}\\
& 15 & 35.9$\pm$1.6 &\textbf{36.9$\pm$1.1}  && 0.87$\pm$0.05  & \textbf{0.90$\pm$0.03}\\
\verb//fastMRI~\cite{zbontar2018fastmri}& 25 & 34.3$\pm$1.4 & \textbf{35.1$\pm$1.1} && 0.85$\pm$0.05 & \textbf{0.89$\pm$0.03}\\
& 35 & 33.1$\pm$1.2 &\textbf{34.1$\pm$1} && 0.82$\pm$0.04 & \textbf{0.88$\pm$0.03}\\
\bottomrule
\end{tabular}
\end{table}

\subsection{Experiments with 3D data with synthetic noise}
Similarly to the 2D experiment, WGN with $\sigma\in[10,15,25,35]$ were added into the test images in the fastMRI~\cite{zbontar2018fastmri} dataset. The quantitative results are tabulated in Table~\ref{table:resultslnee}.
The noise in the simulated BrainWeb images has Rayleigh statistics in the background and Rician statistics in the signal regions. 
The average PSNR (dB) and SSIM results of the BM4D and the proposed ADL on the BrainWeb database are plotted in Figure~\ref{fig:mri3d}. 
ADL was able to reduce the noise effects to a greater extent as compared to BM4D. 
The results of both databases verify the high performance of ADL in noise and RF distortion removal. 
Figure~\ref{fig:visualcomp3d} provides a comparison of the visual results of BM4D and ADL on a noisy T2-weighted image of BrainWab with a noise level of 9 and RF 40\%. 
From the appearance perspective, ADL preserved the histogram better than BM4D. 
Since the edges are blur in the BM4D result, in particular the zoom region, ADL restored the texture successfully. 
In short, the results of 3D are in consonance with 2D results and ADL restored both the appearance and texture information. 

\subsection{Experiments with noisy data with no reference}
\begin{figure*}
    \centering
    \subfigure[Noisy]{
        \centering
        \includegraphics[width=0.25\textwidth, height=0.15\textwidth]{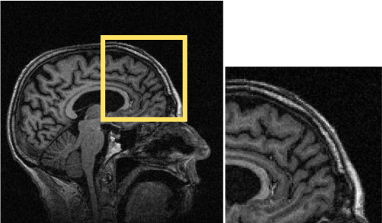}
        \label{fig:norefBrain_noisy}
    }
    \subfigure[Proposed ADL]{
        \centering
        \includegraphics[width=0.25\textwidth, height=0.15\textwidth]{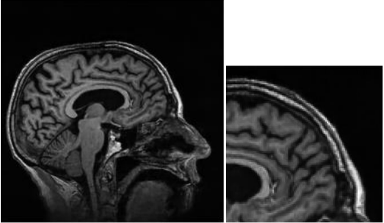}
        \label{fig:norefBrain_adl}
    }
    \subfigure[Histogram]{
        \centering
        \includegraphics[width=0.35\textwidth, height=0.18\textwidth]{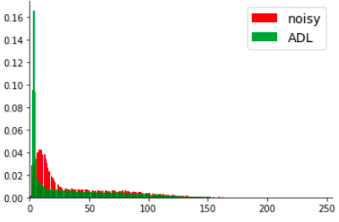}
        \label{fig:norefBrain_hist}
    }
\caption{ADL results on a no-reference noisy Brain MRI
}
\label{fig:norefbrain}
\end{figure*}

\begin{figure*}
    \centering
    \subfigure[Noisy]{
        \centering
        \includegraphics[width=0.25\textwidth, height=0.15\textwidth]{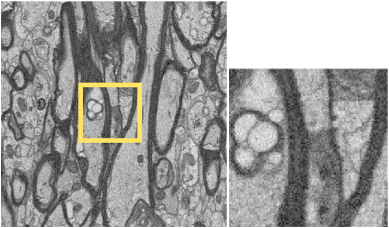}
        \label{fig:norefEM_noisy}
    }
    \subfigure[Proposed ADL]{
        \centering
        \includegraphics[width=0.25\textwidth, height=0.15\textwidth]{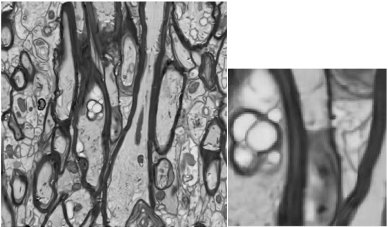}
        \label{fig:norefEM_adl}
    }
    \subfigure[Histogram]{
        \centering
        \includegraphics[width=0.35\textwidth, height=0.18\textwidth]{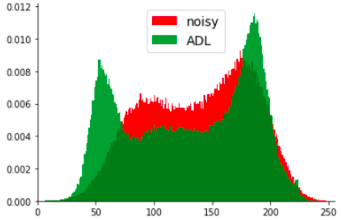}
        \label{fig:norefEM_hist}
    }
\caption{ADL results on a no-reference noisy EM
}
\label{fig:norefem}
\end{figure*}

We experimented with the denoising with a T1-weighted brain MRI from OASIS3-project~\cite{Oasis3}, selected randomly (male, cognitively normal, 87 years), and with an EM dataset from rats' corpus callosum~\cite{abdollahzadeh2021deepacson}. 
The input noisy images, the denoised results by ADL, and the histogram of images are depicted in Figures~\ref{fig:norefbrain} and~\ref{fig:norefem}. 
Both the denoised results and the histogram plots verify that ADL successfully denoised the noisy images and preserved structural information with high precision.

\subsection{Computational complexity} Table~\ref{table::runningTime} reports the running time of the methods in the inference phase. 
The size of data for 2D and 3D experiments is $256\times 256$ and $256\times 256\times 256$, respectively. 
For 2D data, our denoiser needs less than 5 $M$ parameters, much lower than 11.9$M$ of SwinIR. ADL also needs 2$G$ FLOPs, which is considerably lower than SwinIR's. 
For this reason, we named the proposed network as Efficient-UNet. 
For 3D data, our denoiser does not require  heavy computation
of BM4D and 
our method is more appealing for applications requiring immediate responsiveness.

\renewcommand{\tabcolsep}{4pt}
\begin{table}
    \begin{minipage}[h]{1.\linewidth}
    \centering
    \caption{Computational complexity and running time of different methods over 256$\times$256 2D and 256$\times$256$\times$256 3D MRI data.}
    \label{table::runningTime}
    \begin{tabular}{clcccc} 
    \toprule
    &\multirow{2}{*}{Method}
            &{Model Parameters}
                &{FLOPs }
                        &{Running Time }\\
                        &&($M$) & ($G$)& \\
                             \midrule %
    \multirow{2}{*}{2D}&
    \verb//{SwinIR~\cite{liang2021swinir}}&11.9&71.2&539 $ms$\\
    &\verb//{ADL (ours)}&4.75&2.1&143 $ms$\\
    \midrule
    \multirow{2}{*}{3D}&
    \verb//{BM4D}~\cite{maggioni2012nonlocal}&-&-&597.6 $s$ \\
    &\verb//{ADL (ours)}&14.8&121& 14.6 $s$\\
    \bottomrule
    \end{tabular}
    \end{minipage}
    \vspace{0.00mm}
\end{table}

\begin{figure}
\centering
    \includegraphics[width=0.35\textwidth]{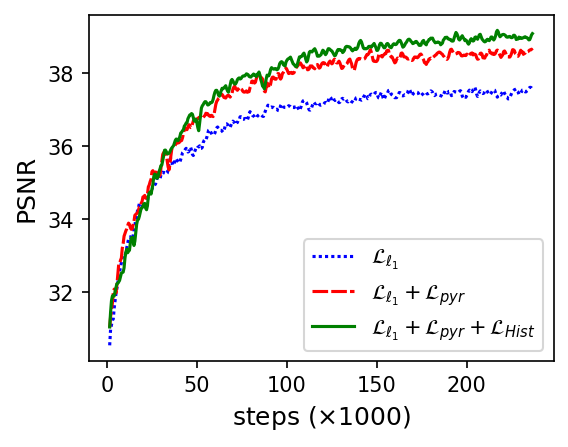}
\caption{PSNR (dB) results of ADL for the validation using different loss functions. The results are averaged over noise levels 10, 15, 25, and 35.}
\label{fig:ablationPSNR}
\end{figure}
\subsection{Ablation study}
\textbf{Multiscale loss functions}: Figure~\ref{fig:ablationPSNR} compares the performance of ADL for different loss functions. The average PSNR results on the validation dataset with $\sigma \in \{10, 15, 25, 35\}$ is reported here. 
It is seen that the pyramidal loss function considerably enhanced PSNR. This scheme was further improved by adding the histogram loss into the training.

\begin{figure}
    \centering
    \includegraphics[width=0.4\textwidth]{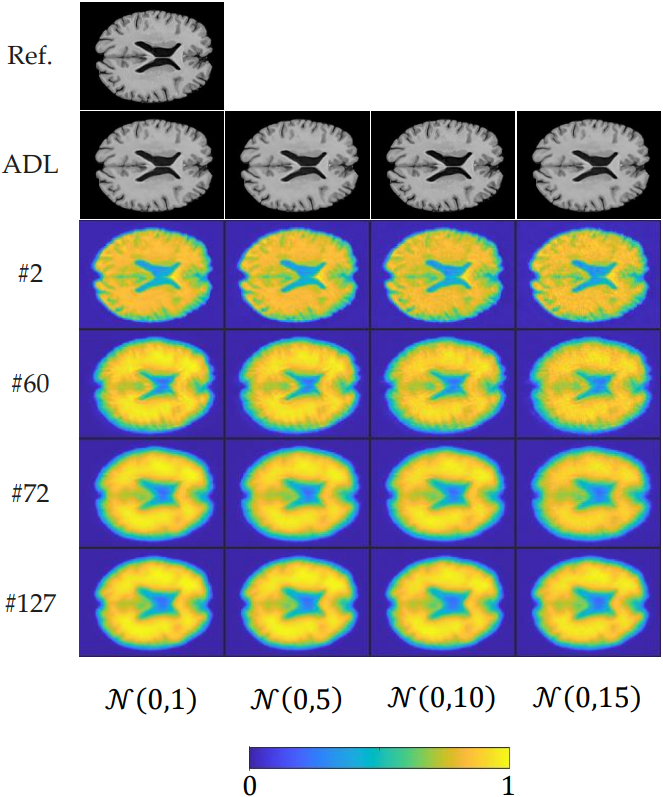}
    \includegraphics[width=0.35\textwidth]{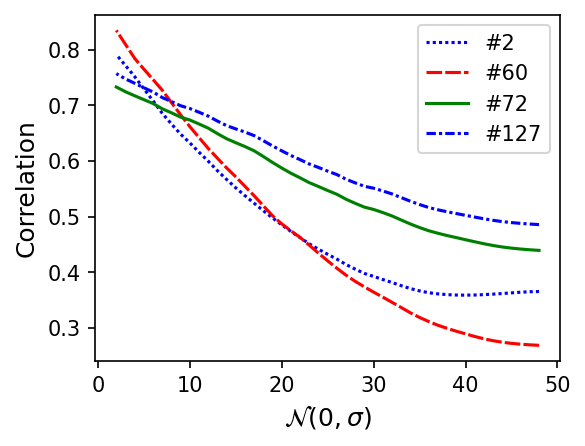}
\caption{Performance of the \textit{Content Enhancer} block in the Efficient-UNet for different noise levels. Left: Visual comparison of the response of a few filters. Bottom: The correlation between the response of the Content Enhancer filters and the proposed ADL outcomes.}
\label{fig:ablationfig}
\end{figure}

\textbf{Content Enhancer block}: As described in Section~\ref{sec:Efficient}, this block contributes to the denoising process directly when the noise level is low. 
In other words, the aim of the Content Enhancer block is to alleviate the vanishing-gradient problem in DNNs when the input noise is low, so it improves the convergence pace. 
We plotted the response of a few filters in the Content Enhancer block in Figure~\ref{fig:ablationfig}. 
We also reported the correlation between the filer responses and the denoised images by ADL. 
It is worth noting that the filter responses for edges and textures are not reported here because it is not straightforward to measure the texture similarity between the edge-oriented filters and the denoised contents. 
When the noise level is below 10, it becomes evident that filters' responses are highly correlated to the denoised contents and this scheme is reduced by increasing the level of AWGN noise. 
It can be referred from the results that the filters of the Content Enhancer block highly contributed to the denoised content when the level of noise is small. 
Since the ADL outcomes remained almost unchanged over high-level noise, we deduce that the network switched to the decoders' filters in this condition.

\section{Conclusion}\label{sec:conclusion}
In this study, we have proposed a novel adversarial distortion learning technique, called ADL, for efficiently restoring images from noisy and distorted observations.
The key idea is to design an auto-encoder called Efficient-UNet that preserves texture and appearance during distortion removal.
Efficient-UNet is based on a hierarchical approach in which low-level features are highly dependent on high-level ones. This enforces the model to restore the images from coarse to fine scales. 
To alleviate the vanishing-gradient problem and to improve the convergence pace, we added the Content Enhancer block to Efficient-UNet.  We proposed a pyramidal loss function for training the ADL to preserve textural information in different scales.
We also introduced a histogram loss to keep the appearance of the denoised contents. 
The proposed model was assessed on 2D and 3D image datasets. Experimental results and a comparative study with the conventional techniques over several 2D and 3D datasets have shown that ADL can restore more accurate denoised contents in the shortest time.

\subsection*{Acknowledgment}
We acknowledge the HPC resources by Bioinformatics Center, University of Eastern Finland. 
This research was partly funded by the Academy of Finland grants \#316258, \#346934  (J. T.) and \#323385 (A. S.).
Data were provided in part by OASIS-3: PIs: T. Benzinger, D. Marcus, J. Morris; NIH P50 AG00561, P30 NS09857781, P01 AG026276, P01 AG003991, R01 AG043434, UL1 TR000448, R01 EB009352.

\bibliographystyle{unsrt}  
\bibliography{references}

\end{document}